\documentclass[10pt,a4paper]{article}
\usepackage{jheppub_kim}
\usepackage{pdflscape}
\usepackage{amsmath}
\usepackage{amssymb}
\usepackage{dcolumn}
\usepackage{bm}
\usepackage{color}
\usepackage{epsfig}
\usepackage{amsfonts}
\usepackage{graphicx}
\usepackage{subfigure}
\usepackage{dcolumn}
\begin{document}
\title{Quantum Thermodynamics of an $\alpha^{\prime}$-Corrected Reissner-Nordstr\"{o}m Black Hole}
\author[a]{Behnam Pourhassan,}
\author[b]{\.{I}zzet Sakall{\i},}
\author[c]{Xiaoping Shi,}
\author[d]{Mir Faizal,}
\author[e]{Salman Sajad Wani}

\affiliation[a] {School of Physics, Damghan University, Damghan, 3671641167, Iran.}
\affiliation[b] {Physics Department, Eastern Mediterranean University, Famagusta 99628, North Cyprus via Mersin 10, Turkey.}
\affiliation[c,d] {Irving K. Barber School of Arts and Sciences, University of British Columbia Okanagan, Kelowna, BC V1V 1V7, Canada.}
\affiliation[d] {Canadian Quantum Research Center 204-3002 32 Ave Vernon, BC V1T 2L7 Canada.}
\affiliation[e] {Department of Physics Engineering, Istanbul Technical University, Istanbul, 34469  Turkey.}

\emailAdd{b.pourhassan@du.ac.ir}
\emailAdd{izzet.sakalli@emu.edu.tr}
\emailAdd{mirfaizalmir@gmail.com}

\abstract{ In this paper, we will analyze the effects of $\alpha^{\prime} $ corrections on the behavior of a Reissner-Nordstr\"{o}m black hole. We will calculate the effects of such corrections on the thermodynamics and thermodynamic stability of such a black hole. We will also derived a novel $\alpha^{\prime}$-corrected first law. We will investigate the effect of such corrections on the Parikh-Wilczek formalism. This will be done using cross entropy and  Kullback-Leibler divergence between the original  probability distribution and the $\alpha^{\prime}$-corrected probability distribution.
We will then analyze the non-equilibrium quantum thermodynamics of this black hole. It will be observed that its quantum thermodynamics is corrected due to  quantum gravitational corrections. We will use Ramsey scheme for emitted particles to calculate the quantum work distribution for this system.
The average  quantum work will be related to the difference of $\alpha^{\prime}$-corrected free energies using the Jarzynski equality.}

\maketitle

\section{Introduction}

The equilibrium thermodynamics of black holes is well suited to describe the behavior of black holes at large distances \cite{1a00,1b,1c,1a0}. In this equilibrium description, temperature of the Hawking radiation is inversely proportional  to its surface gravity, and the entropy of the black hole  scales with its area. So, as the size of the black hole reduces, its temperature increases.
At such short distances, due to high  temperature,   the effect of   thermal fluctuations cannot be neglected. Due to these thermal fluctuations the  equilibrium description breaks down  \cite{32, 32a, 32b, 32c, 32d, 32JHAP}. It is known that  non-equilibrium quantum thermodynamics can be used to describe a   systems where   the equilibrium description   breaks down  \cite{adba0, adba1, adba2, adba4}.
This has motivated the use  of non-equilibrium quantum thermodynamics to describe the behavior of small  black holes \cite{rz12, rz14}.   In the  original Bekenstein-Hawking entropy, a semi-classical approximation is used and  the quantum gravitational corrections are neglected \cite{1a, 2a}. Even though this is well suited for large distance physics of black holes,  it is important to consider quantum gravitationally corrected entropy at  short distances   \cite{6ab, 7ab, 7ba}. As the entropy of black holes is related to the holographic principle \cite{4a, 5a}, these quantum gravitational corrections can also be   holographically obtained using the AdS/CFT correspondence \cite{6a, 7a, 18, 18a, 18b, 18c, 18d}.
The quantum gravitationally corrected thermodynamics for small scales black holes has been studied using string theoretical effects \cite{Dabholkar, ds12, ds14}. It is also possible to use properties of conformal field theory to obtain quantum gravitational corrections to black hole entropy at short distances \cite{2007.15401, Ashtekar, Govindarajan, 29}. The consequences of such   quantum gravitational corrections on the behavior of  black branes has also been studied \cite{b1, b2}. It was observed that at large distances, these quantum gravitational corrections for black branes do not change their thermodynamics, and their effects can be neglected. However, their effects at short distances cannot be neglected, and they produce non-trivial modifications to the thermodynamics of black branes. This is similar to the behavior of black holes \cite{6a, 7a, 18, 18a, 18b, 18c, 18d}.

The non-equilibrium quantum thermodynamics of black solutions modified by quantum gravitational corrections has been studied for small AdS black holes \cite{j1}, Myers-Perry black hole \cite{j2}, and a system of M2-M5 branes \cite{j6}. This quantum work distribution was expressed using  the difference of quantum gravitationally  corrected equilibrium free energies using the Jarzynski equality  \cite{eq12, eq14}. The Jarzynski equality is related to the quantum Crooks fluctuation theorem in non-equilibrium quantum thermodynamics \cite{work1, work2}.
The loss of unitarity in equilibrium black hole thermodynamics has lead to the information paradox \cite{paradox1, paradox2}. However, equilibrium description breaks down at short distances, and we have to use non-equilibrium quantum thermodynamics.
We  observed that the Hawking radiation in  black holes breaks unitarity as it represents heat, and heat in quantum thermodynamics is in turn represented by a non-unitary quantum process \cite{12th, 12tha}. However,  quantum work distribution is represented by a unitary quantum process \cite{12th, 12tha}, so at least there is a unitary quantum process associated with black hole evaporation. Furthermore, this unitary quantum process becomes important at short distances. Thus, it is possible that quantum thermodynamics could potentially resolve the black hole information paradox \cite{j1, j2, j6}.

At short distance it is important to consider quantum gravitational corrections, and $\alpha^{\prime} $ corrections  represent such corrections in string theory.
The $\alpha^{\prime} $ corrections to a black hole   solution have been obtained from the heterotic superstring effective action  \cite{h1}.
These $\alpha^{\prime} $ corrected solutions were used to investigate the effect of these corrections on their equilibrium entropy.  The effect of such corrections on the  entropy  of both BPS and non-BPS
extremal dyonic black holes in heterotic string theory has also been discussed \cite{h2}.
The  near-horizon geometry of such solutions was used to calculate the entropy of these corrected black hole solutions. The   $\alpha^{\prime} $ corrected T-duality transformations have been used to analyze  the invariance of entropy for corrected solutions \cite{h5}.
In this paper, we will analyze the effects of $\alpha^{\prime} $ corrections on non-equilibrium quantum thermodynamics of a Reissner-Nordstr\"{o}m  black hole solution.
The  $\alpha^{\prime} $ corrections to a four dimensional
Reissner-Nordstr\"{o}m  have been explicitly calculated \cite{h1}. This was done using heterotic superstring effective field theory compactified on
 $T^6$.  In fact, using this $\alpha^{\prime} $ corrected metric, the  effect of $\alpha^{\prime} $ corrections on the equilibrium entropy and temperature were also investigated. However, here we will generalize this result  by analyzing the effect of $\alpha^{\prime} $ corrections on the evaporation  of this Reissner-Nordstr\"{o}m   black hole, and  its quantum thermodynamics.

The black hole information paradox has also been addressed by considering the self-gravitation effects of the emitted particles on the black hole geometry    \cite{bb2}. Using this Parikh-Wilczek formalism, it has been argued that the Hawking radiation could be a unitary non-thermal process \cite{bb2}.
In fact,  using the Parikh-Wilczek formalism the black hole entropy has been related to the entropy of such emitted particles \cite{bb1}.
It is  important to analyze the effects of $\alpha^{\prime} $ on  Parikh-Wilczek formalism, as the $\alpha^{\prime} $  can change the behavior of black hole during its last stages. So, in this paper, we will investigate the effect of such quantum gravitational corrections on the  Parikh-Wilczek formalism. We will use   Kullback-Leibler divergence  \cite{kl1, kl2} and cross entropy \cite{c1,c2} to to analyze how the probability density of particles emitted during evaporation is effected by $\alpha^{\prime} $ corrections. The Kullback-Leibler divergence  measures how different two probability distributions are from each other. Furthermore, the Kullback-Leibler divergence  can be related to cross entropy and  entropy of probability distributions. We will use these relations to obtain an expression for the effect of $\alpha^{\prime} $ on such probability distribution. We will also use the Ramsey scheme to calculate the quantum work distribution of these emitted particles \cite{12th, 12tha}. Thus, we will related average quantum work in quantum thermodynamics of black hole  \cite{j1, j2, j6} to the quantum work distribution of emitted particles in the Parikh-Wilczek formalism \cite{bb1, bb2}.

It may be noted that the Parikh-Wilczek formalism has been used to investigate the non-thermal unitary radiation from a  Reissner-Nordstr\"{o}m   black hole \cite{r4}. The thermodynamics of  Reissner-Nordstr\"{o}m black hole is interesting  \cite{r1}, as the  quantum corrections to its  equilibrium thermodynamics can be studied using relativistic quantum geometry \cite{r2}. It has been demonstrated that  small thermal fluctuations to the equilibrium thermodynamics  of  a  Reissner-Nordstr\"{o}m   black hole  can modify its equilibrium entropy \cite{r6}.
It is possible to relate these thermal fluctuations to quantum fluctuations  \cite{32, 32a, 32b, 32c, 32d} using the Jacobson formalism formalism \cite{gr12}. As the geometry   an emergent structure in Jacobson formalism, it is possible to use it to relate the   quantum fluctuation in the geometry to the thermal fluctuations in the equilibrium thermodynamics \cite{gr14}. This formalism has been used to obtain quantum fluctuations of different  black hole solutions \cite{40a, 40b, 40c, 40d}. Thus, the Jacobson formalism \cite{gr12, gr14} can also be used to obtain the quantum corrections to the geometry of a  Reissner-Nordstr\"{o}m   black hole from thermal fluctuations to its equilibrium thermodynamics \cite{r6}. The effect of a zero point length in geometry of spacetime on the thermodynamics of a Reissner-Nordstr\"{o}m   black hole  have been studied using the generalized uncertainty principle \cite{r7}.
The generalized uncertainty principle is a modification to the usual quantum mechanical uncertainty principle \cite{r10}, and can be used to study quantum gravitational corrections to the semi-classical black hole entropy \cite{r12}. The generalized uncertainty principle has been motivated from string theoretical corrections  \cite{r15, r16}. Thus, it is important to analyze  the effects of quantum gravitational corrections on the thermodynamics of a  Reissner-Nordstr\"{o}m  black hole. So, in this paper, we will explicitly analyze such effects produced by the $\alpha^{\prime}$ corrections on such a   Reissner-Nordstr\"{o}m  black hole \cite{h1}.

\section{Quantum Gravitational Corrections}\label{sec1}
\begin{figure}[h!]
 \begin{center}$
 \begin{array}{cccc}
\includegraphics[width=90 mm]{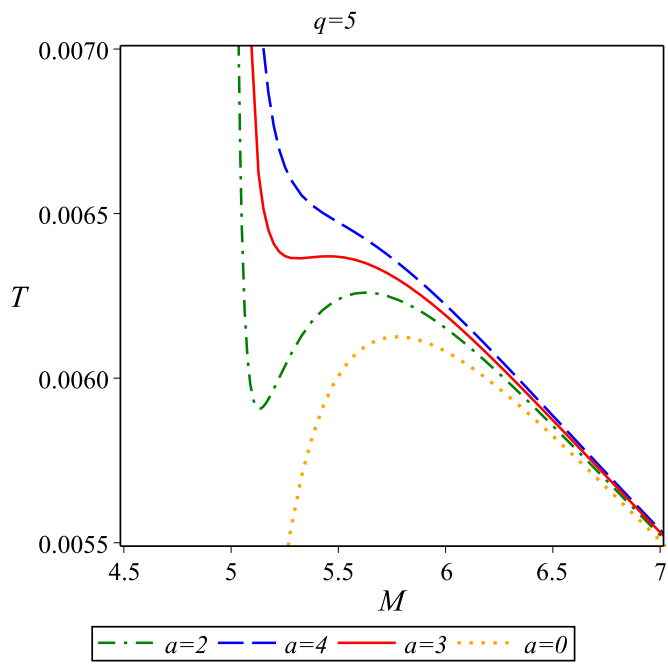}
 \end{array}$
 \end{center}
\caption{Temperature in terms of $M$ with $\alpha^{\prime}=aM^{2}$.}
 \label{fig0}
\end{figure}

It is possible to obtain the $\alpha^{\prime}$ corrections to a four-dimensional Reissner-Nordstr\"{o}m black hole using heterotic superstring effective field theory compactified on
 $T^6$ \cite{h1}. We will analyze the quantum thermodynamics of such a
four dimensional Reissner-Nordstr\"{o}m black hole. Now this solution will be characterized by its  mass $M$ and its electric charge. Furthermore, it will also contain an explicit $\alpha^{\prime}$-corrected contribution. The metric for such a solution can be expressed as \cite{h1},
\begin{equation}\label{metric001}
ds^{2}=N^{2}f(r)dt^{2}-\frac{dr^{2}}{f(r)}-r^{2}(d\theta^{2}+\sin^{2}\theta d\varphi^{2}),
\end{equation}
where
\begin{eqnarray}
N^{2}&=&1+\alpha^{\prime}\frac{p^{2}}{8r^{4}}\nonumber\\
f(r)&=&1-\frac{2M}{r}+\frac{p^{2}}{2r^{2}}-\alpha^{\prime}\frac{p^{2}}{4r^{4}}\left[1-\frac{3M}{2r}+\frac{11p^{2}}{40r^{2}}\right].
\end{eqnarray}
and  $p$ is a physical constant related to the black hole charges. By neglecting $\alpha^{\prime}$ corrections, we obtain the usual uncorrected radius of the event horizon radius \cite{JHAP2},
\begin{equation}
r_{\pm}=M\left(1\pm\sqrt{1-\frac{1}{2}\left(\frac{p}{M}\right)^{2}}\right).
\end{equation}
Now, the $\alpha^{\prime}$-corrected radius of the event horizon is $r_{h}=r_{+}+{\alpha^{\prime}}{M^{-2}}r_{+}^{\prime},$
where $r_{+}^{\prime}$ is defined as \cite{h1}
\begin{equation}
r_{+}^{\prime}\equiv M\left(-\frac{13}{20}\left(\frac{M}{p}\right)^{2}+\frac{8}{5}\left(\frac{M}{p}\right)^{4}
-\frac{\frac{9}{80}-\frac{21}{20}\left(\frac{M}{p}\right)^{2}+\frac{8}{5}\left(\frac{M}{p}\right)^{4}}
{\sqrt{1-\frac{1}{2}\left(\frac{p}{M}\right)^{2}}}\right),
\end{equation}
In the case of single electric charge $q$ it is given by $p=\sqrt{2}q$ \cite{2}. So, we will consider this case of a single charge and
 assume $p=\sqrt{2}q$ for rest of this paper. Here, assuming $ {q}/{M}\ll1$ and $\alpha^{\prime}=aM^{2}$, and so we can write
$
r_{h}=2M+aM( ({M^{2}}/5{q^{2}})- ({9}/{80})).
$
This represents  the quantum gravitationally corrected  radius of the  horizon for this black hole. We can write the corrected temperature of four dimensional Reissner-Nordstr\"{o}m black hole as following \cite{h1},
\begin{equation}\label{Temp}
T=\frac{\gamma}{2\pi M\left(1+\gamma\right)^{2}}
+\frac{\alpha^{\prime}}{M^{2}}\frac{\left(1+3\gamma\right)
\left(1-\gamma\right)^{2}}
{160\pi M\gamma\left(1+\gamma\right)^{5}}.
\end{equation}
Here, for the simplicity we defined $\gamma\equiv\sqrt{1-({q}/{M})^{2}}$.
In Fig. \ref{fig0} we plot the corrected temperature of a four dimensional Reissner-Nordstr\"{o}m black hole as a function of $M$ to analyze its behavior. We observe that for a large  black hole, the $\alpha^{\prime}$ corrections do not change the temperature of the black hole. This is expected as these are quantum gravitational corrections, and should only become important at short distances. We also observe that at intermediate length scales the system becomes very sensitive to value of $\alpha^{\prime}$. However, at very short distances, this sensitive seems on $\alpha^{\prime}$ tends to reduce.

\begin{figure}[h!]
 \begin{center}$
 \begin{array}{cccc}
\includegraphics[width=90 mm]{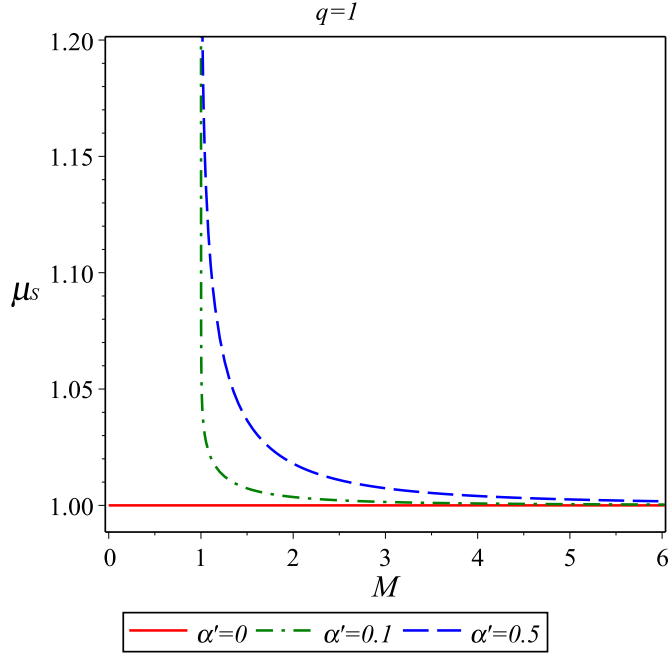}
 \end{array}$
 \end{center}
\caption{$\mu_{S}$ in terms of $M$ with $q=1$.}
 \label{figmus}
\end{figure}

The area of the event  is
$
A_{h}=4\pi r_{h}^{2},
$
while the volume of is $
V= 4 \pi r_{h}^{3}/3.
$. As the entropy of the black hole is related to its area, we can express the entropy of this black hole as  \cite{h1}
\begin{equation}\label{Entropy}
S=\pi M^{2}\left[\left(1+\gamma\right)^{2}+\frac{\alpha^{\prime}}{M^{2}}
\frac{21\gamma^{2}+18\gamma+1}
{40\gamma\left(1+\gamma\right)}\right].
\end{equation}
To analyze the effects of quantum gravitational $\alpha^{\prime} $ corrections on the entropy of this system, we define a $\mu$-coefficient as the ratio of the corrected and original entropies,
  $\mu_{S} = S(\alpha^{\prime})/ S(\alpha^{\prime}=0)$. This $\mu$-coefficient measure the change in entropy due to $\alpha^{\prime}$ corrections
\begin{equation}\label{mu}
\mu_{S}=1+\frac{\alpha^{\prime}}{M^{2}}
\frac{21\gamma^{2}+18\gamma+1}
{40\gamma\left(1+\gamma\right)^{3}}.
\end{equation}
In Fig. \ref{figmus}, we have plotted $\mu_{S}$ for various values of the correction parameter ($\alpha^{\prime}$). We observe that the effect of $\alpha^{\prime}$ corrections reduces as  the size of the black hole increases. This is expected as these corrections are quantum gravitational corrections, there effects can be neglected for large black holes. We also see that the effect produced by $\alpha^{\prime}$ corrections  seems to converge  for small values of mass. This seems to indicate that the effects gravitational corrections will become important at small scales, and different values of $\alpha^{\prime}$ could lead to similar behavior.
We now observe that string tension   can be taken as a variable string theory. In fact, the string tension has been studied in various limits \cite{t1, t2, t3, t4}. It has also been suggested that by a dynamical mechanism can be used to obtain the value of the $\alpha^{\prime}$ \cite{t5, t6}. It has also been proposed that a bound on the $\alpha^{\prime}$ can be obtained from high precision experiments  involving  low energy quantum mechanics systems \cite{t7, t8, t9, t10}. Thus, it is possible to view $\alpha^{\prime}$ as a variable and analyze its effects on thermodynamics. In usual thermodynamics, different state variable their conjugate  are used to analyze a thermodynamic system. This has been done for $S, T$ and $\Phi, q$. In fact, the negative cosmological constant along with its thermodynamic conjugate  has also been used as a state variable \cite{mann1, mann2}. Now if we take the $\alpha^{\prime}$ as a variable, then to specify a state in such  a  system, we have to specify the value of the $\alpha^{\prime}$ along with other thermodynamic variables. Thus, we can view the $\alpha^{\prime}$ as a thermodynamic state  variable, and define $A$ as the conjugate of the $\alpha^{\prime}$, such that
\begin{eqnarray}\label{T-R}
\left(\frac{\partial S}{\partial M}\right)_{q,\alpha^{\prime}}=\frac{1}{T},&&
\left(\frac{\partial S}{\partial q}\right)_{M,\alpha^{\prime}}=-\frac{\Phi}{T},\nonumber\\
\left(\frac{\partial S}{\partial \alpha^{\prime}}\right)_{M,q}=-\frac{A}{T}. &&
\end{eqnarray}
As the entropy is depend on the parameters $M$, $q$ and $\alpha^{\prime}$, we can write a novel $\alpha^{\prime}$-corrected first law of thermodynamics  as
\begin{equation}\label{First-law}
dM=TdS+\Phi dq+Ad\alpha^{\prime}.
\end{equation}
This is the modification to the first law from quantum gravitational corrections.
We obtain the value of the corrected electrostatic potential conjugate to the black hole charge $\Phi$ as
\begin{equation}\label{e-pot}
\Phi=\frac{q}{M\left(1+\gamma\right)}
-\frac{\alpha^{\prime}q^{3}}{80}\frac{8M^{4}\left(1+\gamma\right)
-M^{2}q^{2}\left(11+7\gamma\right)+3q^{4}}{M^{9}
\gamma^{2}\left(1+\gamma\right)^{7}}
\end{equation}
Similarly, by using the $\alpha^{\prime}$ as a state variable, we can obtain an expression for its conjugate $A$. Thus, we can express $A$ as
\begin{eqnarray}\label{A}
A&=&-\frac{160M^{6}\left(1+\gamma\right)-80q^{2}M^{4}\left(5+\gamma\right)
+q^{4}M^{2}\left(321+181\gamma\right)-3q^{6}\left(27+7\gamma\right)}
{80M^{7}\gamma^{2}\left(1+\gamma\right)^{6}}\nonumber\\
&-&\frac{\alpha^{\prime}\left[16M^{4}\left(\gamma-1\right)+4q^{2}M^{2}\left(17-\gamma\right)
+q^{4}\left(63\gamma-51\right)\right]}{{6400M^{7}\gamma^{2}\left(1+\gamma\right)^{6}}}
\end{eqnarray}

\begin{figure}[h!]
 \begin{center}$
 \begin{array}{cccc}
\includegraphics[width=75 mm]{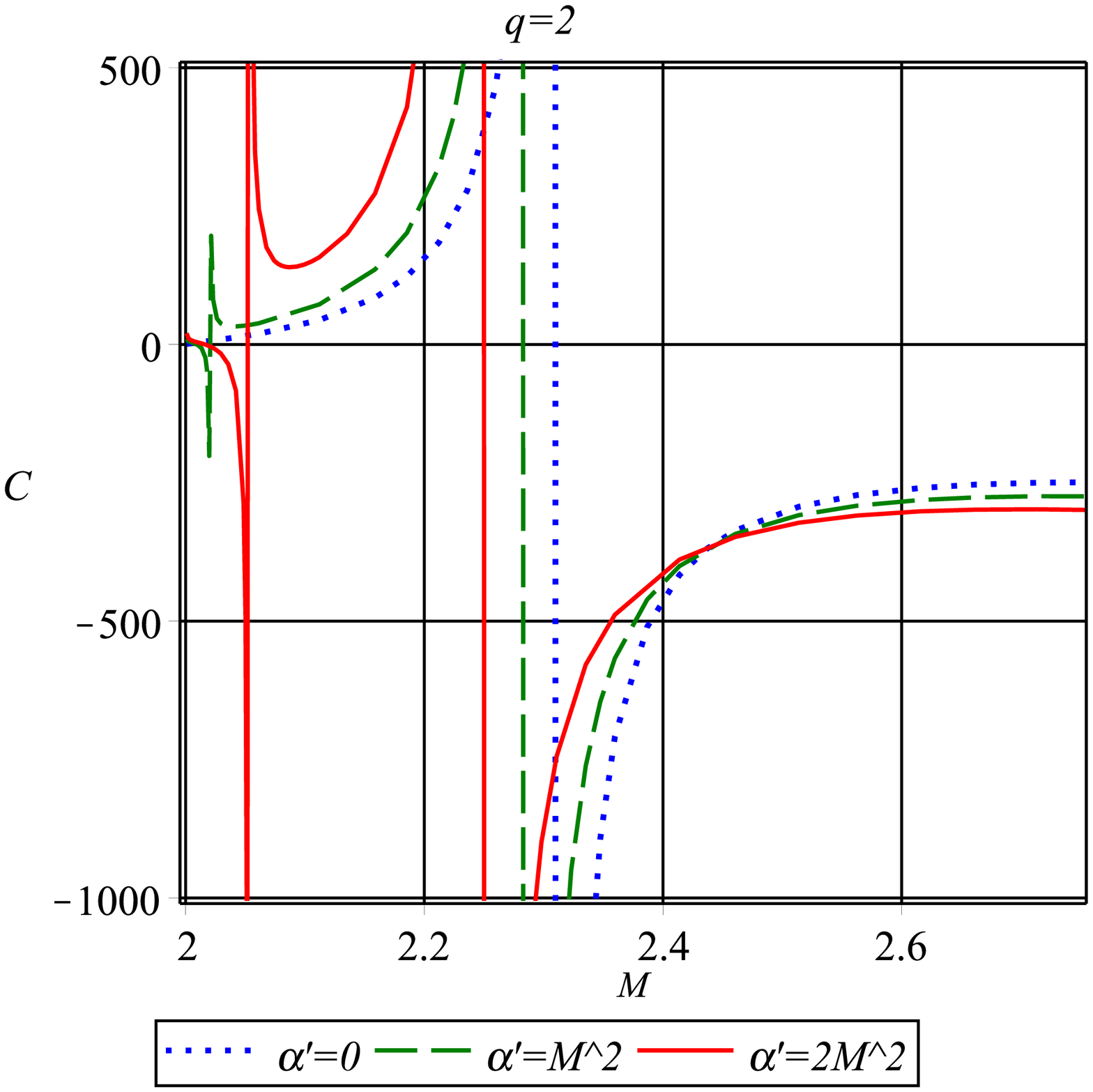}\includegraphics[width=75 mm]{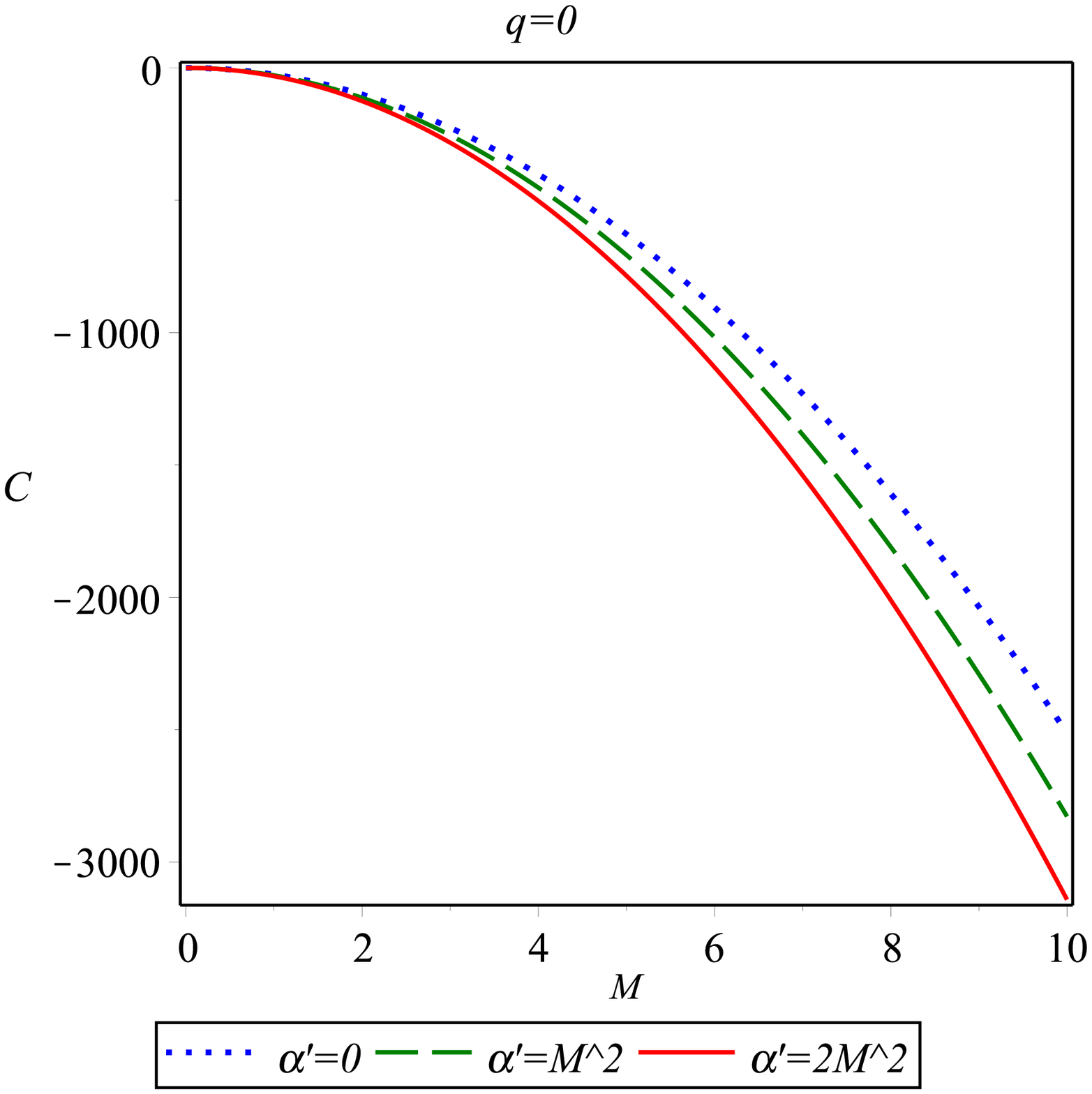}
 \end{array}$
 \end{center}
\caption{Specific heat in terms of $M$ for   $\frac{dT}{dM}\gg\frac{dT}{dq}$.}
 \label{fig1}
\end{figure}

\begin{figure}[h!]
 \begin{center}$
 \begin{array}{cccc}
\includegraphics[width=90 mm]{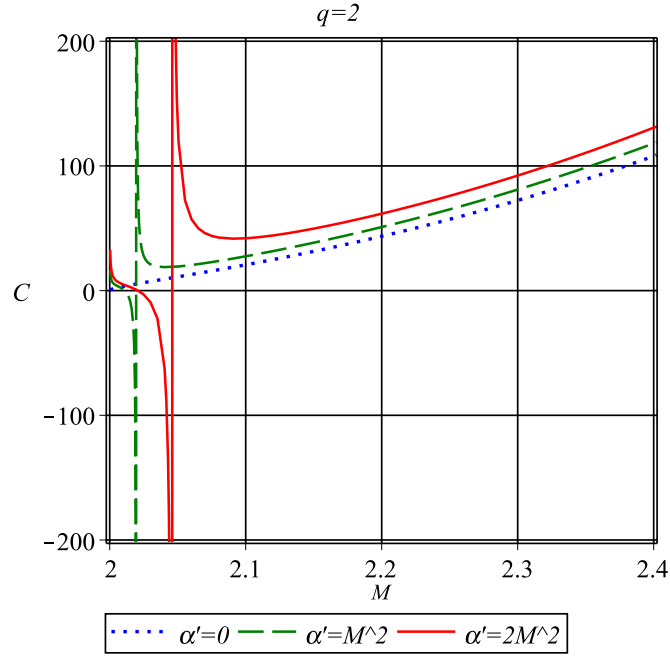}
 \end{array}$
 \end{center}
\caption{Specific heat in terms of $M$ for $\frac{dT}{dM}\ll\frac{dT}{dq}$.}
 \label{fig2}
\end{figure}

Now we can write a quantum gravitationally corrected Smarr-Gibbs-Duhem relation as  ${M}/{2}=TS+({\Phi q}/{2})+A\alpha^{\prime}
$. The entropy and temperature are depend on $M$, $q$ and $\alpha^{\prime}$, hence
\begin{eqnarray}\label{dS}
dS&=&\frac{\partial S}{\partial M}dM+\frac{\partial S}{\partial q}dq+\frac{\partial S}{\partial \alpha^{\prime}}d\alpha^{\prime},
\\
dT&=&\frac{\partial T}{\partial M}dM+\frac{\partial T}{\partial q}dq+\frac{\partial T}{\partial\alpha^{\prime}}d\alpha^{\prime}.
\end{eqnarray}
We observe  that ${dT}/{d\alpha^{\prime}}\ll({dT}/{dM})+({dT}/{dq})$.
So,   for ${dT}/{dM}\gg {dT}/{dq}$, we can write $M\gg q$ and for ${dT}/{dM}\ll {dT}/{dq}$,  we can write  $M=q+\epsilon^{2}$ where $0<\epsilon^{2}<1$ (near extremal case).
Therefore, the specific heat
$
C=T({dS}/{dT}),
$ can be approximated  as,
\begin{eqnarray}\label{C1}
C=\begin{array}{cc}
    T\frac{dS}{dM}\left(\frac{dT}{dM}\right)^{-1}, & \left(\frac{dT}{dM}\gg\frac{dT}{dq}\right)\\
    T\frac{dS}{dq}\left(\frac{dT}{dq}\right)^{-1}, & \left(\frac{dT}{dM}\ll\frac{dT}{dq}\right).
  \end{array}
\end{eqnarray}
Thus, we can express the specific heat as
\begin{equation}\label{C2}
C=-8\pi M^{2}-\frac{5}{2}\pi\left(\frac{q}{M}\right)^{4}\left(M^{2}+\frac{7}{5}q^{2}\right)+\frac{7}{80}\pi\left(\frac{q}{M}\right)^{4}\alpha^{\prime},
  \hspace{1cm} M\gg q.
\end{equation}
By considering zero limit of  $\alpha^{\prime}$ and $q$, we obtain the specific heat for Schwarzschild black hole,  $C=-8\pi M^{2}$. However, near extremal case, we obtain
\begin{eqnarray}\label{C3}
C&=&4\pi^{2}M^{3}-84\sqrt{2}\pi^{2}M^{\frac{5}{2}}\epsilon+504\pi^{2}M^{2}\epsilon^{2}\nonumber\\
&-&\frac{1}{\alpha^{\prime}}\sqrt{2}\pi^{2}M^{\frac{3}{2}}(14720M^{2}+1827\alpha^{\prime})\epsilon^{3}+\mathcal{O}(\epsilon^{4}), \hspace{1cm} M=q+\epsilon^{2}.
\end{eqnarray}
The behavior of the specific heat is plotted in Fig. \ref{fig1} and Fig. \ref{fig2}. In the left panel of Fig. \ref{fig1}, we have plotted  the specific heat of this black hole in terms of the black hole mass for various values of the correction parameter, with first condition (\ref{C1}). The dotted blue line represents the uncorrected case. We observe that there is a phase transition point. It implies that the black hole is in the unstable phase and its mass   reduces as it emits radiation. The dashed green and the solid red lines represent the effect of $\alpha^{\prime}$ correction. Here, we can observe that a second phase transition point occur for  a small mass. This is expected as $\alpha^{\prime}$ correction are a quantum gravitational effect. Therefore, the final stage of the black hole is in an unstable phase. On the other hand from the right panel of Fig. \ref{fig1}, we can observe that the uncharged black hole is completely in an unstable phase.
The specific heat produced by second condition of (\ref{C1}) is plotted in Fig. \ref{fig2}. We can observe that the uncorrected black hole is completely stable, while the corrected black hole has a second order phase transition. This occurs due to the divergence of the specific heat.

\section{Evaporation}
As the black hole evaporates, its entropy will change. We can calculate the change in its entropy, and use it for investigating the effect of the $\alpha^{\prime}$ corrections on the evaporation of such a  black hole. So, one
can express the change in the quantum corrected entropy of the Reissner-Nordstr\"{o}m
black hole as $
\Delta S=S(r_{h2}) - S(r_{h1})
$, where $S(r_{h1})$ is the initial entropy of a $\alpha^{\prime}$-corrected  Reissner-Nordstr\"{o}m
black hole and $S(r_{h2})$ is its final entropy.
Using the relation expression for the $\alpha^{\prime}$-corrected entropy, we can express this change in the entropy during the evaporation process as
\begin{equation}\label{Delta-Entropy}
\frac{\Delta S}{\pi}= M_{2}^{2}\left[\left(1+\gamma_{2}\right)^{2}+\frac{\alpha^{\prime}}{M_{2}^{2}}
\frac{21\gamma_{2}^{2}+18\gamma_{2}+1}
{40\gamma_{2}\left(1+\gamma_{2}\right)}\right]
-M_{1}^{2}\left[\left(1+\gamma_{1}\right)^{2}+\frac{\alpha^{\prime}}{M_{1}^{2}}
\frac{21\gamma_{1}^{2}+18\gamma_{1}+1}
{40\gamma_{1}\left(1+\gamma_{1}\right)}\right],
\end{equation}
Here $\gamma_1$  is   expressed in terms of initial mass and  charge  $(M_1, q_1)$. Similarly,  $\gamma_2$ is  expressed in terms of  final mass and charge   $(M_2, q_2)$. Thus, we can write
\begin{eqnarray}\label{gamma-1-2}
\gamma_{1}=\sqrt{1-\left(\frac{q_{1}}{M_{1}}\right)^{2}}, &&
\gamma_{2}= \sqrt{1-\left(\frac{q_{2}}{M_{2}}\right)^{2}}
\end{eqnarray}

It has been argued that the Hawking radiation might be a unitary non-thermal process, when the self-gravitational effects of
the emitted particles on the black hole geometry are not neglected    \cite{bb2}.
In fact,  using this Parikh-Wilczek formalism \cite{bb2}    the black hole entropy has been can be related to the   entropy of such particles emitted  \cite{bb1}. Thus, we can relate  the change in the entropy of the $\alpha^{\prime}$-corrected black hole to the entropy of such particles, using the $\alpha^{\prime}$-corrected version of the  Parikh-Wilczek formalism.
Even though certain problems remain in this approach \cite{bb4, bb5}, this formalism can be used to analyze the effect of the $\alpha^{\prime}$ correction on information in a black hole.  Using Parikh-Wilczek formalism \cite{bb2}, we first denote  the number of ways   in
which the original classical black hole can   evaporated by emission of  $i$ particles as $\Omega_i (M, q)$. Now the total number of ways in which such a classical black hole will evaporate will be $ \Omega_t (M, q) = \sum_{i =1}^\infty \Omega_i (M, q)$. So, the probability that the black hole will  evaporated by the emission of $i$ particles is given by $Q_i = \Omega_i (M, q)/ \Omega_t (M, q)$.
Similarly, we can write the $\alpha^{\prime}$-corrected probability as $P_i(\alpha^{\prime}) = \Omega_i (M, q;\alpha^{\prime} )/ \Omega_t (M, q; \alpha^{\prime} )$, where $\Omega_i (M, q)$ and $ \Omega_t (M, q)$ are quantum gravitationally corrected $\Omega_i (M, q)$ and $ \Omega_t (M, q)$. Here, the quantum gravitational corrections are taken up to the order $\alpha^{\prime}$.  Thus, we observe that  $Q_i = P_i(\alpha^{\prime} ) |_{\alpha^{\prime}  =0}$. Using Parikh-Wilczek formalism \cite{bb1, bb2},  entropy corresponding to the original and $\alpha^{\prime}$-corrected  distributions  can be represented by $ S[Q]$ and $S[P]$, such that \cite{bb1, bb2}
\begin{eqnarray}
S[Q] = -\sum_{i =1}^\infty\,  Q_i \log Q_i, &&
S[P] = -\sum_{i =1}^\infty\, P_i(\alpha^{\prime} ) \log P_i(\alpha^{\prime} )
\end{eqnarray}
We can also write the cross entropy between the original and $\alpha^{\prime}$-corrected probability distributions as  \cite{c1, c2}
\begin{equation}
H [P, Q] = - \sum_{i =1}^\infty\, P_i(\alpha^{\prime} ) \log Q_i
\end{equation}
As  we are using the $\alpha^{\prime} $ as a state variable,  we can analyze how the probability distribution changes due to its variation. Since $\alpha^{\prime} $ is small, we can use the Taylor expansion of $P_i(\alpha^{\prime} )$ near $\alpha^{\prime} =0$ and $Q_i = P_i (\alpha^{\prime} )|_{\alpha^{\prime}  =0}$ to obtain
\begin{eqnarray}
   H[P, Q]&=&-\sum_{i =1}^\infty\,  \left[P_i(\alpha^{\prime} )+\alpha^{\prime}   \frac{\delta P_i(\alpha^{\prime} )}{\delta \alpha^{\prime} }\right]|_{\alpha^{\prime} =0} \log Q_i  \nonumber
\\&=&   S[Q] - \alpha^{\prime}  \sum_{i =1}^\infty\,  \left[ \frac{\delta P_i(\alpha^{\prime} )}{\delta \alpha^{\prime} }\right]|_{\alpha^{\prime} =0} \log Q_i
\end{eqnarray}
Similarly, we can also use the Taylor expansion of $P_i(\alpha^{\prime} )$ near $\alpha^{\prime} =0$ and  write $S[P]$ as
\begin{eqnarray}
S[P] &=&   S[Q]   - \alpha^{\prime}  \sum_{i =1}^\infty\, \left[   \frac{\delta P_i(\alpha^{\prime} )}{\delta \alpha^{\prime} }\right]|_{\alpha^{\prime} =0}  \log Q_i -  \alpha^{\prime}  \sum_{i =1}^\infty\,\left[   \frac{\delta P_i(\alpha^{\prime} )}{\delta \alpha^{\prime} }\right]|_{\alpha^{\prime} =0}
\end{eqnarray}
The entropy and cross entropy can be used to obtain  Kullback-Leibler divergence as
$
   D_{KL}(P||Q)= H[P, Q] - S[P]
$, and so we  can  write the Kullback-Leibler divergence for these probability distributions \cite{kl1, kl2}
\begin{equation}
D_{KL}(P||Q)= \sum_{i =1}^\infty\,  P_i(\alpha^{\prime} ) \log \frac{P_i(\alpha^{\prime} )}{Q_i} = \alpha^{\prime}  \sum_{i =1}^\infty\,\left[   \frac{\delta P_i(\alpha^{\prime} )}{\delta \alpha^{\prime} }\right]|_{\alpha^{\prime} =0}
\end{equation}
It   measure   how different the original probability  distribution  $Q_i$  is   from the $\alpha^{\prime}$-corrected probability distribution  $P_i(\alpha^{\prime} )$.
So, using the expansion for  entropy and cross entropy for the small $\alpha^{\prime}$, we observe that the Kullback-Leibler divergence  scales with $\alpha^{\prime}$. Thus, we observe  that larger values $\alpha^{\prime}$ would produce larger difference between the probability distribution of particles emitted in Parikh-Wilczek formalism \cite{bb1, bb2} and the $\alpha^{\prime} $-corrected probability distribution. Even though the Kullback-Leibler divergence is not a statistical distance, as it is not symmetric, it does measure how different two distributions are from each other. Hence, it was was used to analyze the  effect of the $\alpha^{\prime}$ corrections on the probability distributions of emitted particles.

\section{Quantum Work Distribution}\label{sec4}
We will analyze the quantum work distribution for an evaporating     black hole. The quantum work distribution has been analyzed for various different  black hole solutions \cite{j1, j2, j6}. However, as quantum work becomes important at short distance scales, it is important to explicitly analyze the effects of quantum gravitational corrections on quantum work. Thus,  we will analyze the effects of the $\alpha^{\prime}$ corrections on quantum work distribution. We first observe that the internal energy of a  Reissner-Nordstr\"{o}m black hole can be calculated from its entropy  \cite{Do11}. Thus, a change in the entropy of  a black hole will also change its internal energy. We can use the $\alpha^{\prime}$-corrected Reissner-Nordstr\"{o}m entropy to calculate the change in its internal energy as it evaporates. The internal energy of the $\alpha^{\prime}$-corrected Reissner-Nordstr\"{o}m can be written as
\begin{equation}\label{In.E}
E=\frac{40 r_{h}^{14} \left(19 X_{1}-8 X_{2}\right)
+3168 q^8 \alpha^{\prime3} \left(8192 X_{1}-X_{2}\right)}{15 r_{h}^5 \left(r_{h}^4-224 \alpha^{\prime}  q^2\right) \left(16 r_{h}^4-6 \alpha^{\prime}  q^2\right) X_{1}}
\end{equation}
where we defined,
$X_{1}= \sqrt{{2\alpha^{\prime}  q^2}{r_{h}^{-4}}+8}, $ and $
X_{2}=\sqrt{128 \alpha^{\prime}  q^2{r_{h}^{-4}}+2}$.
Now we can obtain  the  change in the internal energy of this black hole as it evaporates from an initial radius   $r_{h1}$ to the  the final radius is $r_{h2}$ as $
\Delta E=E(r_{h2}) - E(r_{h1})$.
 At large scales, the change in the internal energy occurs mostly because of   radiation.
However, as the black hole becomes sufficient small, the average   quantum work becomes important. This average   quantum work is also     an unitary   process in quantum thermodynamics \cite{12th, 12tha}.
 So, apart  from the Parikh-Wilczek formalism \cite{bb1, bb2},   a part of the energy will still be lost through another   unitary quantum process represented by quantum work distribution \cite{j1, j2, j6}. Now   we  denote this  average  quantum work done  by $\langle W \rangle$, and the  energy lost through non-thermal radiation in the Parikh-Wilczek formalism  as $Q$  as  this black hole evaporates from $r_{h1}$ to $r_{h2}$,\cite{bb1, bb2}. The change in the internal energy can only occur due to these two processes, and so we can write $\Delta E =  Q - \langle W \rangle$ \cite{12th, 12tha}.

\begin{figure}[h!]
 \begin{center}$
 \begin{array}{cccc}
\includegraphics[width=65 mm]{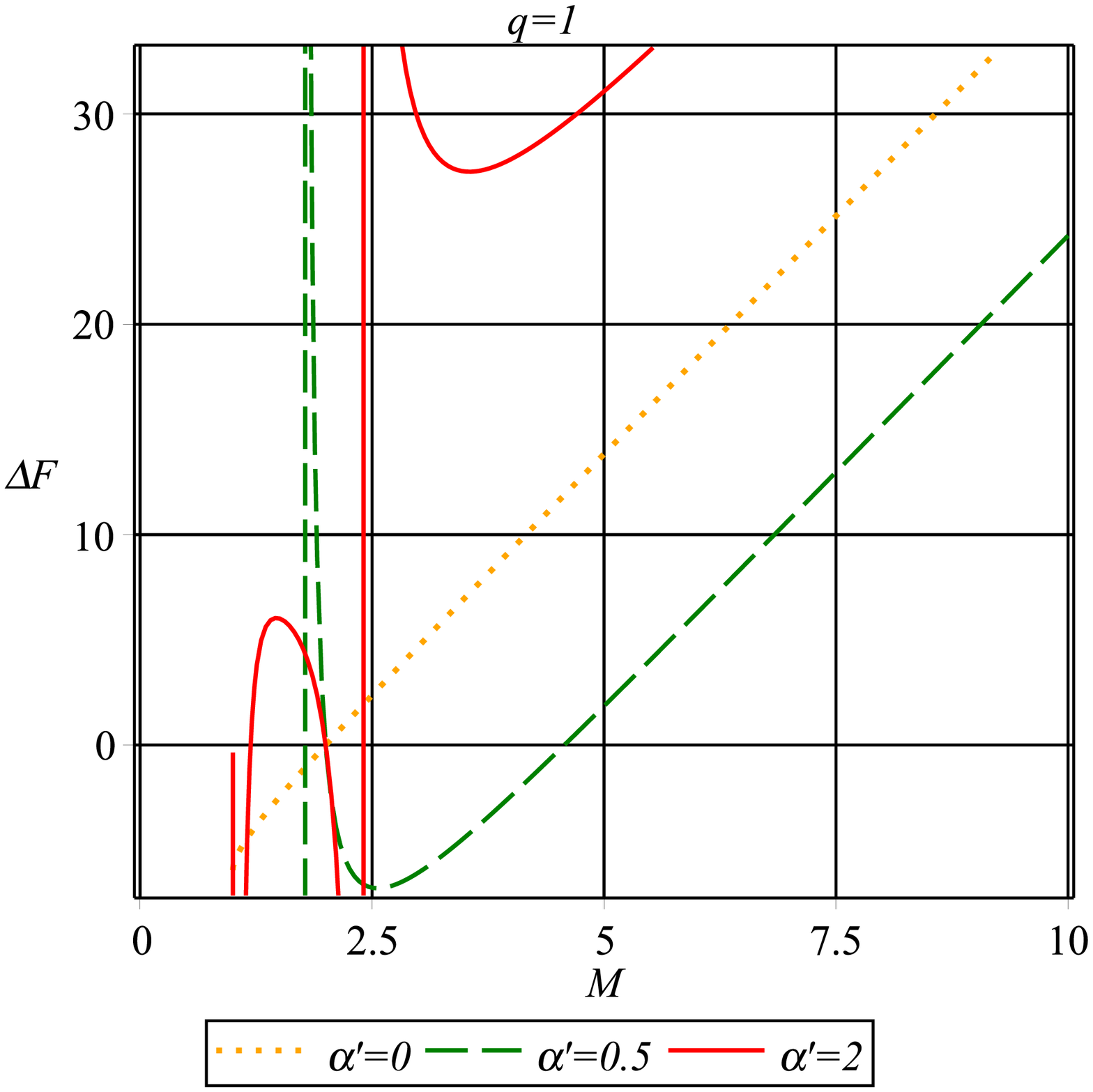}\includegraphics[width=65 mm]{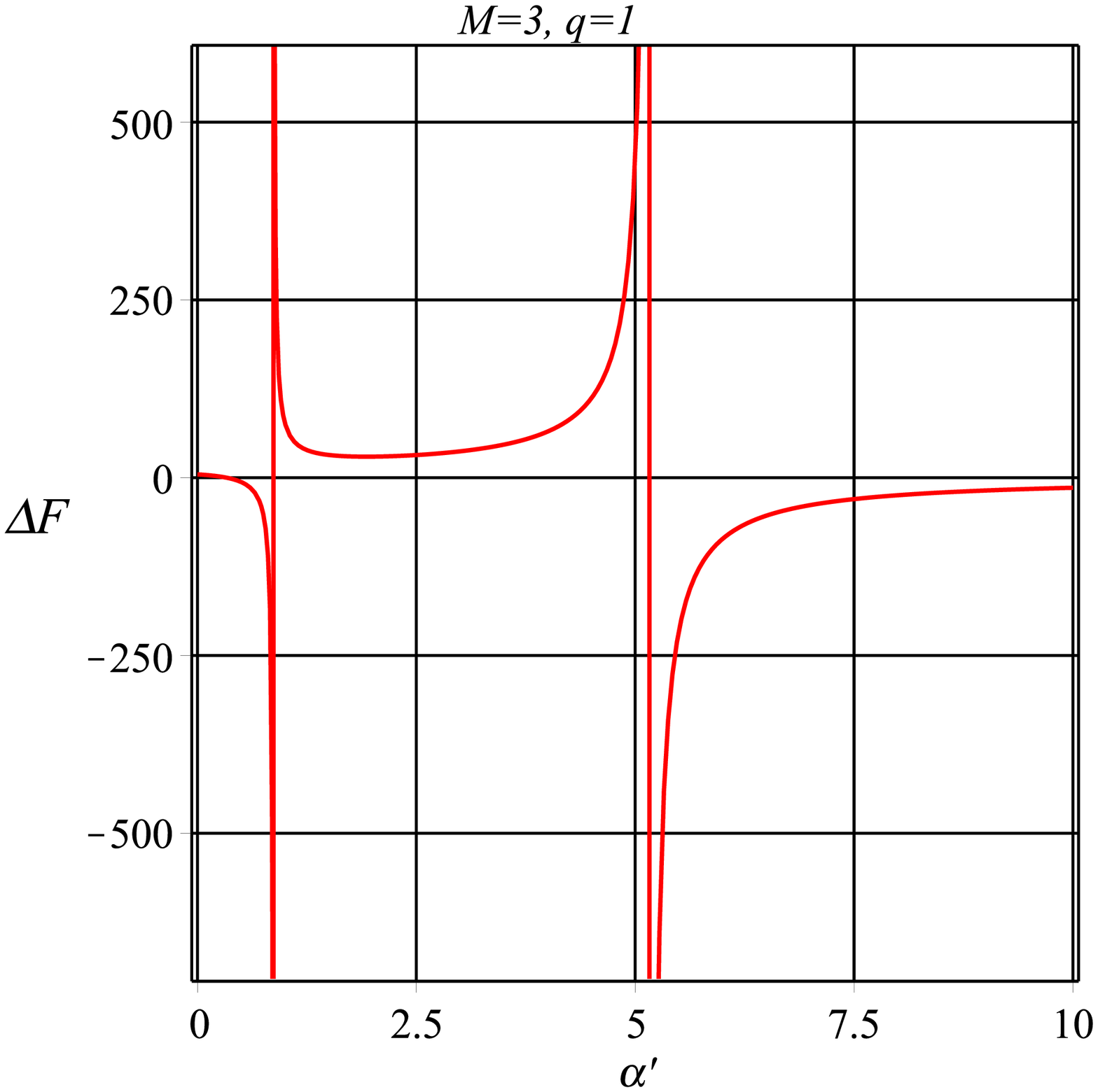}
 \end{array}$
 \end{center}
\caption{Left: $\Delta F$ in terms of $M$ for $q_{1}=1$ and $M_{1}=2$. Right: $\Delta F$ in terms of the $\alpha^{\prime}$ for $q_{1}=1$ and $M_{1}=2$.}
 \label{fig-Delta-F}
\end{figure}

We can explicitly analyze the average quantum work for the emitted particles.
As the black hole evaporates, the energy of the particles emitted will also change.
If the initial  energy eigenvectors are $|E_i\rangle$ and the final energy eigenvectors are $|{\tilde E}_j\rangle$, then we can calculate quantum work distribution by measuring the change in the Hamiltonian  \cite{work1, work2}. However, as the particles are described by relativistic field theories, we have to use  Ramsey scheme to obtain quantum work distribution \cite{12th, 12tha}.
Now if $\rho$ is the density matrix for the emitted particles, then the  probability for getting energy    $E_i$ at time $\tau =0$  is  $p^0_i = |\langle {E}_i|\rho|{E}_i\rangle|^2$. The   eigenvalues   of Hamiltonian evolves from $|E_i\rangle $ to $|\tilde E_j\rangle$  as the black hole evaporates. In Parikh-Wilczek formalism \cite{bb1, bb2} this evolution is represented by a unitary operator  $\hat{U}$.
The   conditional probability to obtain energy    $\tilde E_l$ at  $\tau = t$, if the initial energy was $E_i$ at   $\tau =0$ can be represented by  $p^{\tau}_{l|i} = \langle \tilde {E}_j|U | {E}_i\rangle|^2 $. The difference between the  initial and the final energies is given by $
\Bar{W}_{i,j}= \tilde E_j-E_i.
 $
The probability associated with the occurrence  of this difference $\Bar{W}^{i,l}$ can be obtained  from $ p^0_i $ and $ p^{\tau}_{l|i}$ as
  \begin{eqnarray}
p_{i,j} &=& p^0_i p^{\tau}_{j|i}
= |\langle {E}_i|\rho|{E}_i\rangle |^2|\langle \tilde {E}_j|U | {E}_i\rangle|^2
\end{eqnarray}
Now we represent quantum work  distribution by the variable ${W}$. In absence of   degeneracies,  this would coincide with  $\Bar{W}_{i,l}$. The   probability distribution associated with $W$ can    be expressed as
$
P({W})=\sum_{ij}p_{i,j}\delta(\mathcal{W}-\Bar{W}_{i,j}).
$ Using this  probability distribution, and the expression for   $p_{i,j}$, the average work done can be written as
  \begin{eqnarray}
          \langle W \rangle = \int \sum_{ij}p_{i,j}\delta(W-\Bar{W}_{i,j})W dW
              \langle  W \rangle
         = \sum_{ij} |\langle {W}_i|\rho|{W}_i\rangle|\langle \tilde {W}_j|U | {W}_i\rangle|^2 \left[\tilde W_j (\tau) -W_i (0)\right]
     \end{eqnarray}
It is possible to express this average quantum work as $ \langle  W \rangle  = tr[\hat{H}{(\tau)}\rho{(\tau)}]-tr[\hat{H}{(0)}\rho{(0)}]$ \cite{12th, 12tha}. The  characteristic function for quantum work distribution   $W$ can be expressed as
  \begin{equation}
    \Tilde{P}(\mu)=\int P(W)e^{i\mu W} dW=\langle e^{i\mu W}\rangle
  \end{equation}
where   $\mu$ is  a parameter which is used   in the Ramsey scheme. Here, we start from an auxiliary qubit, with   $|0\rangle$ as its ground state,  and $|1\rangle$ as its   excited state.   We first couple this   auxiliary qubit in the ground state  to the
 density state $\rho (0)$ of emitted particles at time $t =0$, and write
$
\hat{\rho}_\text{tot}=   \hat{\rho}\otimes\hat{\rho}_\text{aux}.
$
Next we apply a Hadamard operator to the qubit.
After the system has evolved from $ \hat{H}(0)$ to $\hat{H}(t)$ due to the evaporation for $t$,   we can write   $
     \hat{C} (\mu) = \hat{U} e^{-i\mu  \hat{H}(0) } \otimes|0\rangle\langle0|+ e^{-i\mu  \hat{H}(t)}\hat{U}\otimes|1\rangle\langle1|
$. The qubit state can  be written as $
    \hat{\rho}_\text{aux}= \text{Tr} _{X}[\hat{C}(\mu) \hat{\rho}_\text{tot} \hat{C}^\dagger (\mu)] $,
where    $\text{Tr}_{X}$, represents a trace over all states of emitted particles.  Now we apply a   second  Hadamard operation to the qubit state. Using this   Hadamard operation we have extracted the information about the evolution  of the system due to the evaporation of the black hole. As we have an expression for $\hat{\rho}_\text{aux}  $,  it can be  compared to the general expression for   $\hat{\rho}_\text{aux}   $, which is given by
$2 \hat{\rho}_\text{aux}    =
  1 + \text{Re} [\tilde{P}(\mu )]\hat{\sigma}_z + \text{Im} [\tilde{P}(\mu )]\hat{\sigma}_y ~ $ \cite{12oth, 12otha}. Using this expression we obtain an expression for the  characteristic function $\tilde{P}(\mu )$. This characteristic function can then be used to obtain the average quantum work  as
\begin{eqnarray}
\langle W\rangle  =i  \frac{d}{d\nu}
\Tilde{P}(\nu).
\end{eqnarray}
The Jarzynski equality can be used to calculate this average quantum work done as this black hole evaporates between two states \cite{work1, work2}. This is done by relating this average quantum work to the difference of the equilibrium free energies for black hole \cite{eq12, eq14}
\begin{equation}
\langle e^{-\beta W} \rangle = e^{\beta \Delta F}
\end{equation}
Here, we have used the Jensen equality  to relate  the average of the exponential to the exponential of the average  as
$ \exp {\langle -\beta {W} \rangle } \leq \langle \exp {(-\beta W)} \rangle
$. The equilibrium free energy of the $\alpha^{\prime}$-corrected black hole can be written  as
\begin{eqnarray}\label{Hel.Energy}
F&=& \Big[{960 r_{h}^5 \left(r_{h}^4-224 \alpha^{\prime}  q^2\right) \left(16 r_{h}^4-6 \alpha^{\prime}  q^2\right) X_{1}}\Big]^{-1}
\nonumber \\ &\times&\Big[
-2560 r_{h}^{14} \left(3 \sqrt{2}-19 X_{1}+8 X_{2}\right)+2560 q^2 r_{h}^{12} \left(3 \sqrt{2}-253 X_{1}+8 X_{2}\right)\nonumber\\
&+&960 \alpha^{\prime} q^2 r_{h}^{10} \left(1795 \sqrt{2} -5908  X_{1}+2056   X_{2}\right)-64 \alpha^{\prime}  q^4 r_{h}^8  \left(26910 \sqrt{2}-632287 X_{1}+30800 X_{2}\right)\nonumber\\
&-&160 \alpha^{\prime2}  q^4 r_{h}^6 \left(4041 \sqrt{2} -3160320  X_{1}+4632  X_{2}\right)\nonumber\\
&+&24 \alpha^{\prime2} q^6 r_{h}^4  \left(17953 \sqrt{2}-185185024 X_{1}+20568 X_{2}\right)\nonumber\\
&+&46080 \alpha^{\prime3} q^6 r_{h}^2 \left(7 \sqrt{2}  -4096   X_{1}+8   X_{2}\right)-50688 \alpha^{\prime3} q^8 \left(7 \sqrt{2}-65536 X_{1}+8 X_{2}\right) \Big]
\end{eqnarray}
Using this expression for the equilibrium   free energy, we can calculate the difference between free energies as $\Delta F=F(r_{h2}) - F(r_{h1})$.
In Fig. \ref{fig-Delta-F}, we have plotted the effect of the  $\alpha^{\prime}$ on $\Delta F$. We observe that  $\alpha^{\prime}$ modify free energies and thus average quantum work for a  Reissner-Nordstr\"{o}m black hole.
It is possible to construct a microscopic model for a stretched horizon of the Reissner-Nordstr\"{o}m black hole, where the  horizon consists of discrete constituents \cite{so12}. Using this model it is possible to write an expression for the partition function for this black hole. The black hole partition function can  also been constructed using string theory \cite{so14}.
 As the black hole evaporates, the  partition function  associated with the black hole will  change. As the black hole evaporates   from an initial partition function $Z_1$    to a final partition function $Z_2$,   the relative weight of these partition functions  $Z_2/Z_1$ can be related to the average quantum work as  $\langle \exp {(-\beta W)} \rangle ={Z_2}/{Z_1}$ \cite{eq12, eq14}.
Furthermore, using the relation between average quantum work and equilibrium free energies, this   relative weights of the partition function can also be related to  the difference between the  equilibrium  free energies of the black hole as
$\exp {\beta \Delta F} = {Z_2}/{Z_1}
$. Thus, we can obtain information information about the behavior of this black hole, and its $\alpha^{\prime}$ correction using average quantum work.

\section{Conclusion}\label{Con}
We have studied the effects of $\alpha^{\prime} $ corrections on the behavior of Reissner-Nordstr\"{o}m black hole.
The metric for such black holes had already been constructed using  heterotic superstring effective field theory compactified on
 $T^6$ \cite{h1}.
We have observed that the $\alpha^{\prime} $ corrections change  thermodynamic stability of  a  Reissner-Nordstr\"{o}m black hole. As the thermodynamical state will depend on the value of  $\alpha^{\prime} $, we have used it as a thermodynamic state variable and defined a conjugate variable corresponding to it. We have obtained a novel  form of the first law using this $\alpha^{\prime} $ correction.
We have  used Parikh-Wilczek formalism to investigated the effect of $\alpha^{\prime} $ corrections on non-thermal radiation.
This was done using the  Kullback-Leibler divergence  and cross entropy for the original  probability distribution and the $\alpha^{\prime} $ corrected  probability  distribution.  The  non-equilibrium quantum thermodynamics for this $\alpha^{\prime} $ corrected    black hole was also investigated.  We have  used Ramsey scheme for emitted particles to calculate the quantum work distribution for this system.
The Jarzynski equality  was used to  relate average  $\alpha^{\prime} $ corrected average quantum work to the difference of $\alpha^{\prime} $ corrected free energies. This was done by using non-equilibrium quantum thermodynamics for this black hole. As we have used $\alpha^{\prime} $ corrected solutions, we were able to study the   effect of $\alpha^{\prime} $ correction on this quantum thermodynamics of this black hole.

It would be interesting to analyze higher order $\alpha^{\prime} $ corrections for such black holes. We can investigate the effects of  higher order $\alpha^{\prime} $ corrections on the Parikh-Wilczek formalism. This can be used to analyze how higher order quantum gravitational corrections change the    non-thermal radiation. We can also relate these quantum gravitationally corrected probability distributions, for each other of  $\alpha^{\prime} $  to each other. This can be done using the Kullback-Leibler divergence  and cross entropy for the each of those   probability distributions.   We can also investigate the effect of such  higher order $\alpha^{\prime} $ corrections on  non-equilibrium quantum thermodynamics for this $\alpha^{\prime} $. We can again use Jarzynski equality to investigate  the effects of $\alpha^{\prime} $ corrections on the average quantum work.
Furthermore, $\alpha^{\prime} $ corrections can be obtained for various different black hole solutions, and these solutions can then be used to investigate the effects of such corrections on the quantum thermodynamics of very small black holes. As the quantum work is a unitary process it would be interesting to investigate the effect of such corrections on quantum work distribution of different black hole solutions. This can then be related to information  paradox. It would be possible to study such $\alpha^{\prime} $ corrections for AdS black holes, and then use the Jarzynski equality  to obtain average quantum work  for such corrected AdS solutions. Furthermore, as we have constructed a novel modification  of the first law, it would be interesting to combine the modification proposed in this paper, with extended phase space thermodynamics of an AdS black hole.  We can also relate the behavior of an AdS space to its field theoretical dual. Thus, we can analyze a field theoretical dual to $\alpha^{\prime} $  corrected quantum work distribution  for an AdS black hole.

\end{document}